\documentclass[10,5 pt, a4paper]{article}%\documentclass[10,5 pt, a4paper]{article}
\usepackage{epsfig}
\usepackage{epsf}
\usepackage{bm}                 % for bold math symbols
\usepackage{amsmath}
\usepackage{amssymb}
\usepackage{dcolumn}
\usepackage{graphicx}
\usepackage{psfrag}
\usepackage{latexsym}
\usepackage{color}

\newcommand{\beq}{\begin{equation}}
\newcommand{\eeq}{\end{equation}}
\newcommand{\beqa}{\begin{eqnarray}}
\newcommand{\eeqa}{\end{eqnarray}}

\def\Symp#1,#2,#3,#4.{\left[\left(\begin{array}{c}#1\\#2\end{array}\right),\left(\begin{array}{c}#3\\#4\end{array}\right)\right]}
\def\Vec#1,#2.{\left(\!\begin{array}{c}#1\\#2\end{array}\!\right)}
\def\vec#1,#2.{{#1\choose{#2}}}

\newcommand{\bx}{{\bf x}}
\newcommand{\by}{{\bf y}}

\definecolor{redcom}{rgb}{1,0.1,0.2}
\definecolor{querycol}{rgb}{0.2,0.2,1}

\definecolor{purplerep}{rgb}{1,0.1,1}

\definecolor{green}{rgb}{0.1,0.8,1}

\begin{document}

\title{Generalized guidance equation for peaked quantum solitons and effective gravity.}
\date{}
\author{}
\maketitle
\centerline{Thomas Durt\footnote{ Aix  Marseille  Universit\'e,  CNRS,
Centrale  Marseille, Institut Fresnel (UMR 7249),13013 Marseille, France.email: thomas.durt@centrale-marseille.fr}}

% Force line breaks with \\

\abstract{Bouncing oil droplets have been shown to follow de Broglie-Bohm like trajectories and at the same time they exhibit attractive and repulsive pseudo-gravitation. We propose a model aimed at rendering account of these phenomenological observations. It inspires, in a more speculative approach, a toy model for quantum gravity.}
\maketitle
In studies concerning quantum wave mechanics, de Broglie-Bohm (dB-B) trajectories \cite{debrogliebook,bohm521,Holland} remained during many years a rather confidential and academic topic, but they regained interest since they were realized in the lab. with artificial macroscopic systems, the bouncing oil droplets or walkers  \cite{Busch,fort}, which were shown experimentally to follow dB-B like quantum trajectories.  For instance, when the walker passes through one slit of a two-slit device, it undergoes the influence of its ``pilot-wave'' passing through the other slit, in such a way that, after averaging over many dB-B like trajectories, the interference pattern typical of a double-slit experiment is restored, despite of the fact that each walker passes through only one slit. Pseudo-gravitational interaction has also been reported between two droplets. In \cite{fort} for instance we can read: {\it ...We find that, depending on the value of d}, (d represents here the impact parameter of the collision){\it the interaction is either repulsive or attractive. When repulsive, the drops follow two approximately hyperbolic trajectories. When attractive, there is usually a mutual capture of the two walkers into an orbital motion similar to that of twin stars  ...}

Our study was motivated by the double solution program \cite{1927,debrogliebook,debroglieend} of Louis de Broglie, according to which the quantum wave function, solution of the linear Schr\"odinger equation does not contain all elements of reality concerning the quantum system. In de Broglie's view, particles would consist of a peaked concentration of energy that would coexist with the linear wave function, and the latter would guide the former according to the de Broglie-Bohm guidance equation. Our work consists of an attempt to realize de Broglie's program by adding another ingredient to it, which is that we suppose the existence of a non-linear self focusing potential  (\ref{selfclas}) of gravitational nature which prohibits the spreading of the peaked soliton (solitary wave) representing the particle. Therefore we assume that the system evolves in time according to the non-linear equation (\ref{notfreeNL}).

\begin{equation}
{i}\hbar\frac{\partial\Psi(t,{{\bf x}})}{\partial t}=-\hbar^2\frac{\Delta\Psi(t,{{\bf x}})}{2m}
+V^{L}(t,{{\bf x}})\Psi(t,{{\bf x}})+V^{NL}(\Psi)\Psi(t,{{\bf x}}),\label{notfreeNL}
\end{equation}
where $V^{L}(t,{{\bf x}})$ represents the usual external potentials, for instance it could represent the coupling to external electro-magnetic fields. It is thus represented by a self-adjoint operator linearily acting on the Hilbert space. $V_{NL}$ represents a self focusing non-linearity. Contrary to $V^{L}$ it depends on $\Psi$. 
In order to distinguish short range and long range we shall consider in the rest of the paper the Yukawa like screened S-N potential:
\beqa V^{NL}(\Psi)\equiv -K\int {d}^3 x'(\frac{|\Psi(t,{{\bf x'}})|^2}{|{{\bf x} -{\bf x'}}|})e^{-{|{{\bf x} -{\bf x'}}|\over \lambda}},\label{selfclas}\eeqa 
where $K$ is a positive coupling constant and $\lambda$ is a characteristic length to be defined soon. The Schr\"odinger-Newton (S-N) potential is in turn defined as follows: $V_{SN}^{NL}(\Psi)((t,{{\bf x}}) \equiv -Gm^2\int {d}^3 x'(\frac{|\Psi(t,{{\bf x'}})|^2}{|{{\bf x} -{\bf x'}}|})$; (\ref{notfreeNL}) has been abundantly studied in the literature in presence of the S-N potential and in absence of external potential ($V^L=0$). 
%Then (\ref{notfreeNL})  reads
%\begin{equation}
%{i}\hbar\frac{\partial\Psi(t,{{\bf x}})}{\partial t}=-\hbar^2\frac{\Delta\Psi(t,{{\bf x}})}{2m}+V_{SN}^{NL}(\Psi)\Psi(t,{{\bf x}}),\label{freeNL}\end{equation} 
In particular, when  $V^L=0$, (\ref{notfreeNL}) possesses a ``ground-state" solution of the form
$\psi(\bx, t) = {e}^{-\frac{{i}E_g t}{\hbar}} \phi_{NLg}(\bx)$; $\phi_{NLg}$ is thus  a solution of 

\beqa\frac{-\hbar^2}{2m} \Delta \phi_{NLg}(\bx) + G m^2\! \int\!d^3y ~\!\frac{\big| \phi_{NLg}(\by)\big|^2}{|\bx-\by|}~\! \phi_{NLg}(\bx) \nonumber\\= E_g  \phi_{NLg}(\bx)~\!,\label{Choquard}
\eeqa
which was studied in astrophysics and is known under the name of the {\em Choquard} equation \cite{Lieb,CDW}.  In \cite{Lieb}, Lieb showed that the energy functional 
\beqa
E(\phi) = \frac{\hbar^2}{2 m} \int\!{d}^3x~\big|\nabla\phi(\bx)\big|^2\label{Choquard-energy} \\ \nonumber
- \frac{G m^2}{2} \int d^3x \int d^3y\!\frac{~\big|\phi(\by)\big|^2}{|\bx-\by|}~\!\big|\phi(\bx)\big|^2~\!,
\eeqa
is minimized by a unique solution $ \phi_{NLg}(\bx)$ of the Choquard equation (\ref{Choquard}) for a given $L_2$ norm $N( \phi)$.

As is well-known, the Choquard equation possesses well-defined scaling properties \cite{CDW,vanMeter}; in particular, we may scale its solution in such a way that in the limit where the $L_2$ norm goes to plus infinity, the extent goes to zero, and the energy to minus infinity. Numerical studies also established that $ \phi_{NLg}$ has a ``hump'', quasi-gaussian shape \cite{CDW,vanMeter}. 
In the case of quantum particles the non-linear potential $V^{NL}$ aims represents in our eyes a short-range self-focusing interaction of gravitational nature which localizes the wave function  $\Psi$, analogous to the Poincar\'e pressure \cite{Poincar} aimed at ensuring the cohesion of electrons in presence of Coulombian self-repulsion. In the rest of the paper we shall identify the quantum particles with these self-collapsed wave packets (solitons).
Our aim is to show that when several particles are present there appears a long range effective gravitational interaction between them. We explicitly established a distinction between gravitation at short and long spatial scale in our model,  with the aim of gaining information about the short scale behaviour by fitting the long range effective gravitational potential with Newton gravitation. 
In the case of bouncing droplets, the non-linearity mimicks the complicate mechanism which ensures their stability. That we consider droplets or quantum systems, $\lambda$ is intentionally chosen to be larger than the size of the object under interest (particle or droplet), in such a way that the solitonic ground state of (\ref{notfreeNL}) when $V^L=0$ is essentially the solution $ \phi_{NLg}(\bx)$ of (\ref{Choquard}). If we consider quantum particles like the electron or the proton, we choose $K$ to be $Gm^2$; in the case of droplets $K$ ought to be calibrated in function of the experimental set-up.

% and $\lambda$ to be the Planck length (more or less $10^{-35}$ meter), which is quite larger than their size as we shall show. If we consider bouncing oil droplets, we modelize the , making use of (\ref{selfclas}) and imposing then $\lambda$ to be, say, ten times the size of the typical size of the contact surface between the droplet and the liquid on which it bounces. 
%In a previous paper \cite{old,new}, we solved a non-linear evolution equation by imposing (\ref{ansatzspin},\ref{S1V},\ref{S2bis}) that the wave function of the system could be factorized (\ref{ansatzspin}) into the product of a smoothly varying function, solution of the usual linear Schr\"odinger equation (\ref{S1V}), with a peaked soliton, of which the spread is countered by a self-focusing non-linearity (\ref{S2bis}). This constitutes the first non-standard ingredient of our approach.

The first real novelty of our paper is that we assume that the solution of (\ref{notfreeNL}) obeys the factorisation ansatz:

 \beqa \Psi=({1\over A_L(t,{\bf x_0(t)})})\Psi_L(t,{\bf x})\cdot \phi_{NLg}({\bf x}-{\bf x_0}(t))e^{-iE_g t/\hbar},\label{approsingle}\eeqa
where we introduce the pilot wave $\Psi_L$, of amplitude $A_L$ and phase $\varphi_L$: $\Psi_L(t,{\bf x}))\equiv A_L(t,{\bf x})e^{i\varphi(t,{\bf x})}$ and ${{\bf x_0}}$ the position of the barycentre of the soliton $\phi_{NLg}$. We also suppose to begin with that ${{\bf x_0}}$ follows a de Broglie-Bohm trajectory: \beqa \label{bari} {{\bf x_0}}(t)={{\bf x_0}}(t=0)+\int_0^t dt {\bf v}_{dB-B},\eeqa where ${\bf v}_{dB-B}$ represents the dB-B velocity derived from the pilot wave $\Psi_L$ evaluated at the position of the barycentre of $\Psi$:
 
 \beqa{\bf v}_{dB-B}={\hbar \over m}{{\bf \bigtriangledown}}\varphi_L({{\bf x_0}}(t),t).\label{dBB}\eeqa
 
 %In \cite{old} we explain how we were brought to this choice(\ref{selfclas},\ref{approsingle},\ref{bari},\ref{dBB}), but in the present paper we treat it as a postulate for reasons of convenience. Actually in \cite{old} we studied the feedback of the pilot wave $\Psi_L$ on the soliton or particle $\phi_{NL}$ and we argued that this feedback results in conditions (\ref{approsingle},\ref{bari},\ref{dBB}) in the appropriate regime.  
 
Our goal is to study the feedback of the soliton on the pilot wave. We shall thus study in the rest of the paper solutions of (\ref{notfreeNL},\ref{selfclas}), under the constraints (\ref{approsingle},\ref{bari},\ref{dBB}).

The second novelty of our paper is that contrary to the majority of the papers that can be found in the literature concerning the S-N equation (\ref{notfreeNL}), we assume, for reasons that we shall clarify in the core of the paper, that whenever we consider elementary particles the size of the solution of  (\ref{notfreeNL},\ref{approsingle}) is very small, actually of the order of $Gm/c^2$. In the case of an electron for instance this size is of the order of 10$^{-55}$ meter. This means \cite{CDW} that the energy (\ref{Choquard-energy}) of the self-collapsed ground state is of the order of $-mc^2$. In our approach, the external potentials are thus treated as a perturbation, contrary to the standard approach, where the non-linearity is treated as a weak perturbation \cite{smolin,arxiv}.  The non-linearity is so strong in our case that it forces the state of the system to behave as a solitary wave, a soliton of extremely small size, solution of the so-called Choquard equation (\ref{Choquard}) for which the spread gets compensated by the non-linearity. The wave function of the system is thus supposedly self-collapsed to begin with, since arbitrary long times. This marks a difference with previous studies which rather focused on the collapse process itself \cite{diosi84,penrose,vanMeter,CDW,arxiv}. We consider situations where $A_L$ and $\varphi_L$ smoothly  vary in space and time compared to $\phi_{NLg}$ so that in good approximation (\ref{approsingle}) reads  \beqa \Psi =e^{i\varphi_L}\cdot \phi_{NLg}({\bf x}-{\bf x_0}(t))e^{-iE_g t/\hbar},\label{approsinglebis}\eeqa
which explicitly shows that $\Psi$ is a constant L$_2$ norm function, in accordance with the norm-preserving character of the evolution equation (\ref{notfreeNL}).
In this paper, we shall apply these ideas to a system consisting of two particles in order to derive an effective long range gravitational attraction between the particles. We also compare our predictions to droplets phenomenology and make in this case new predictions that are likely to be tested in the lab.

%We also discuss the possibility to test our predictions in bouncing oil droplets \cite{Bush,fort,fort2,fort3} phenomenology, where similar effects (dB-B trajectories and effective gravitation) have already been observed in the past. We explicitly show how our model provides a simple explanation of the reported observations of repulsive gravitation \cite{fort}, as well as of the quantization of twin-star-like orbiting motion in the case of attractive gravitation \cite{fort3}. 

%We actually consider that the implication of our model is twofold: it seemingly provides a good model for simulating the properties of droplets, but it could also have more fundamental implications.

\section{Effective gravitation}

To simplify the treatment we shall first consider distinguishable particles $A$ and $B$ and suppose that the wave function of the full system reads

$\Psi({\bf x},t)= \Psi_L ({\bf x},t)\cdot ({\phi_{NLg}^{A}({\bf x},t)\over A_L({\bf x^A_0},t)}+{\phi_{NLg}^{B} ({\bf x},t)\over A_L({\bf x^B_0},t)})$
where we introduced the short notation 

$\phi_{NLg}^{A(B)} ({\bf x},t)\equiv \phi_{NLg}({\bf x}-{\bf x^{A(B)}_0}(t))e^{-iE_g t/\hbar},$ with ${\bf x_0^A}$ and ${\bf x_0^B}$ the positions of the barycentres of the solitons assigned to the two particles.
In virtue of (\ref{approsinglebis}), at the locations ${{\bf x^{A(B)}_0}}$, $V^{NL}(\Psi)=V^{NL}(\phi_{NLg})$, so that $\phi_{NLg}^{A(B)} ({\bf x},t)$ satisfies the constraint

%the non-linear component $\phi_{NL}(t,{{\bf x}})\equiv ({\phi_{NLg}^{A}({\bf x},t)\over A_L({\bf x^A_0},t)}+{\phi_{NLg}^{B} ({\bf x},t)\over A_L({\bf x^B_0},t)})$ 
 \beqa &&i\hbar\cdot  \frac{\partial\phi_{NLg}^{A(B)}}{\partial t}=\nonumber
-\frac{\hbar^2}{2m}\cdot \Delta\phi_{NLg}^{A(B)}+V^{NL}(\Psi)\phi_{NLg}^{A(B)} \label{aux}\\&&-\frac{\hbar^2}{m}\cdot i{\bf \bigtriangledown}  \varphi_L(t,{{\bf x^{A(B)}_0}}) \cdot {\bf \bigtriangledown} \phi_{NLg}^{A(B)} \label{const}\eeqa

Multiplying (\ref{const}) by $ \Psi_L ({\bf x},t)$, and substracting the result from (\ref{notfreeNL}) we find, in the low-dBB-velocity regime where the temporal derivative of $A_L({\bf x^{A(B)}_0},t)$  is small compared to other contributions  \beqa && ({\phi_{NLg}^{A}({\bf x},t)\over A_L({\bf x^A_0},t)}+{\phi_{NLg}^{B} ({\bf x},t)\over A_L({\bf x^B_0},t)})({i}\hbar\cdot  \frac{\partial \Psi_L(t,{\bf x})}{\partial t}+\nonumber\\ &&
 \frac{\hbar^2}{2m}\Delta\Psi_L(t,{{\bf x}})-V^{L}(t,{{\bf x}})\Psi_L(t,{{\bf x}}))= \nonumber\\ 
 && -\frac{\hbar^2}{m}({\bf \bigtriangledown}   \Psi_L(t,{{\bf x}}) + \Psi_L(t,{\bf x})\cdot i{\bf \bigtriangledown}  \varphi_L(t,{{\bf x_0}}))\cdot  \nonumber \\     && {\bf \bigtriangledown} ({\phi_{NLg}^{A}({\bf x},t)\over A_L({\bf x^A_0},t)}+{\phi_{NLg}^{B} ({\bf x},t)\over A_L({\bf x^B_0},t)})                 .\label{DS1a}\eeqa

 We shall only consider situations where  the positions ${\bf x^A_0}$ and ${\bf x^B_0}$ are not too close to each other e.g. larger than $\lambda$; then we may neglect the overlap between the two solitons, so that, dividing (\ref{DS1a}) by $({\phi_{NLg}^{A}({\bf x},t)\over A_L({\bf x^A_0},t)}+{\phi_{NLg}^{B} ({\bf x},t)\over A_L({\bf x^B_0},t)})$ we find

  \beqa &&{i}\hbar\cdot  \frac{\partial \Psi_L(t,{\bf x})}{\partial t}+
 \frac{\hbar^2}{2m}\Delta\Psi_L(t,{{\bf x}})-V^{L}(t,{{\bf x}})\Psi_L(t,{{\bf x}})= \nonumber\\ 
 && (-\frac{\hbar^2}{m}({\bf \bigtriangledown}   \Psi_L(t,{{\bf x}}) + \Psi_L(t,{\bf x})\cdot i{\bf \bigtriangledown}  \varphi_L(t,{{\bf x_0}}))\cdot  \nonumber \\     && ({{\bf \bigtriangledown} \phi_{NLg}^{A}({\bf x},t)\over \phi_{NLg}^{A}({\bf x},t)}+ {{\bf \bigtriangledown} \phi_{NLg}^{B}({\bf x},t)\over \phi_{NLg}^{B}({\bf x},t)}        ),\label{DS1b}\eeqa
which looks like the linear Schr\"odinger equation with source terms. At some distance from the soliton $\phi_{NLg}$ decreases exponentially, because $\phi_{NLg}$ is a negative energy state with radial symmetry.
Moreover ${{\bf \bigtriangledown}_{x,y,z} \phi_{NLg}({\bf x},t)\over \phi_{NLg}({\bf x},t)} $ is odd in $x,y,z$ around $x_0,y_0,z_0$. It behaves thus along the $x,y,z$ axis passing through $(x_0,y_0,z_0)$ as a kink with a s-like shape $_{---}$/$ ^{---}$. In the plane orthogonal to this axis its tail decreases like $(x-x_0)/\sqrt{(x-x_0)^2+(y-y_0)^2+(z-z_0)^2}$, but we should not worry about the tails because when several particles are isotropically distributed, the tails cancel each other and the shape of the kink is rather $---/---$. The integration over space of the source term proportional to $\Psi_L$ is equal to zero. Integrating the source term proportional to ${\bf \bigtriangledown}   \Psi_L$ by part we find that 
everything happens as if the source terms in equation (\ref{DS1a}) were proportional to Dirac 3-deltas, and we may write (\ref{DS1a}) in the form

\beqa &&{i}\hbar\cdot  \frac{\partial \Psi_L(t,{\bf x})}{\partial t}+\frac{\hbar^2}{2m}\Delta\Psi_L(t,{{\bf x}}) -V^{L}(t,{{\bf x}})\Psi_L(t,{{\bf x}})\nonumber \\&&= \label{DSKb}
-4\pi\frac{\hbar^2}{2m} \Psi_L(t,{{\bf x}}) \cdot   \\ \nonumber&&(L_A\delta^{Dirac}({\bf x}-{\bf x_0^A})+L_B\delta^{Dirac}({\bf x}-{\bf x_0^B})),\eeqa
%$+\int \int dy dz f(t,{{\bf x}}) \frac{{\bf \bigtriangledown}_x\phi_{NL}(t,{{\bf x}}{ \phi_{NL}(t,{{\bf x}})}$
where  $L$ is a length of the order of the size (extent) of the particle. We incorporated in $L_{A(B)}$ factors of the order of unity, in order to simplify the computations.

In a perturbative or bootstrap approach, we shall now treat  (\ref{DSKb}) as a generalised Poisson equation with a source term proportional to a solution of the homogeneous (linear Schr\"odinger) equation without sources; moreover we impose a factorisable solution in the form 

$\Psi^{inhom}_L(t,{{\bf x}})=\Psi^{hom}_L(t,{{\bf x}})(-\phi^G(t,{{\bf x}})/c^2)$, then (\ref{DSKb}) reads

\beqa \nonumber &&(-\phi^G(t,{{\bf x}})/c^2)({i}\hbar\cdot  \frac{\partial \Psi^{hom}_L(t,{\bf x})}{\partial t}+\frac{\hbar^2}{2m}\Delta\Psi^{hom}_L(t,{{\bf x}})  \nonumber\\ 
&&\nonumber -V^{L}(t,{{\bf x}}))+\frac{\hbar^2}{m} {\bf \bigtriangledown}\Psi^{hom}_L(t,{{\bf x}})\cdot  {\bf \bigtriangledown} (-\phi^G(t,{{\bf x}})/c^2)\\
&&\nonumber +(\Psi^{hom}_L(t,{\bf x}))({i}\hbar\cdot  \frac{\partial (-\phi^G(t,{{\bf x}})/c^2)}{\partial t}+\\\nonumber
  &&\frac{\hbar^2}{2m}\Delta(-\phi^G(t,{{\bf x}})/c^2))=-4\pi\frac{\hbar^2}{2m} \cdot   \Psi^{hom}_L(t,{{\bf x}})\cdot \\ && (L_A\delta^{Dirac}({\bf x}-{\bf x_0^A})+L_B\delta^{Dirac}({\bf x}-{\bf x_0^B})).\label{bigb}\eeqa

The three first terms of (\ref{bigb}) are equal to zero, because $\Psi^{hom}_L$ is solution of the linear Schr\"odinger equation without sources. It is consistent, in the low-dBB-velocity regime, to neglect 

$ {\bf \bigtriangledown}\Psi^{hom}_L(t,{{\bf x}})\cdot  {\bf \bigtriangledown} (-\phi^G(t,{{\bf x}})/c^2)$ as well as $\frac{\partial (-\phi^G(t,{{\bf x}})/c^2)}{\partial t}$ as we shall explain soon. Then $\phi^G$ obeys the Poisson equation
\beqa \Delta(\phi^G(t,{{\bf x}})/c^2)=\nonumber \\ 4\pi\cdot(L_A\delta^{Dirac}({\bf x}-{\bf x_0^A})+L_B\delta^{Dirac}({\bf x}-{\bf x_0^B}))\label{fishb}\eeqa
Making use of the well-known properties of the Green functions associated to the Laplace equation, it is easy to check that the solution of (\ref{fishb}) is

\beqa \phi^G(t,{{\bf x}})/c^2=-\left({L_A\over |{{\bf x}}-{{\bf x_0^A}}|}+{L_B\over |{{\bf x}}-{{\bf x_0^B}}|}\right)\label{symb}.\eeqa
 $ \phi^G(t,{{\bf x}})/c^2$ looks thus like an effective (here repulsive) gravitational potential\footnote{Note that the singularities of this expression in ${{\bf x_0^A}}$ and ${{\bf x_0^B}}$ are artificial, they result from our discretisation procedure. }. We check a posteriori that we were in right to neglect certain corrections proportional to the dB-B velocities.

 %In \cite{new}, we have shown, by estimating the generalised Hamilton-Jacobi equation of the system that everything happens as if an effective gravitational potential is present, resulting from the perturbative scheme outlined here. Here we shall rederive this result in a slightly different manner.

In order to fit $\phi^G$ with a Newtonian potential in the long range domain, let us consider $H_L$ the linear Hamiltonian of the system:

$H_L=H_L^A+H_L^B+m_Ac^2+m_Bc^2+V_L^{AB}$, including a possible coupling between the $A$ and $B$ systems. In the non-relativistic regime, $H^{A(B)}_L<<m_{A(B)}c^2$, $H_L^{AB}<<m_{A(B)}c^2$, $|{i}\hbar\cdot  \frac{\partial \Psi}{\partial t}|<<m_{A(B)}c^2|\Psi_L|$.

Note that $ \Psi _L(t,{{\bf x}})$, the solution of (\ref{DSKb}) obeys $\Psi_L$ =$\Psi_L^{hom}+\Psi^{inhom}=(1-\phi^G(t,{{\bf x}})/c^2)\Psi_L^{hom}$. Let us from now on neglect self-interactions which were already taken account in $V^{NL}$. Requiring that $\Psi^{hom}_L =   \Psi _L /(1-\phi^G/c^2)$ satisfies the homogeneous equation $H_L\Psi^{hom}_L ={i}\hbar\cdot  \frac{\partial \Psi^{hom}_L }{\partial t}$ results, provided we only consider the dominating contribution, in the replacement in $H^{A(B)}$ of $m_{A(B)}c^2$ by $m_{A(B)}c^2(m_{A(B)}c^2)\phi_{A(B)}^G(t,{{\bf x}})/c^2$. At this level, an effective gravitational interaction between $A$ and $B$ emerges which satisfies $\phi_{A(B)}^G(t,{{\bf x}})/c^2=-({L_{B(A)}\over |{{\bf x_0^{B(A)}}}-{{\bf x^{A(B)}_0}}|})$.

 Requiring the equivalence of this effective potential with  the Newtonian gravitational potential $\phi^{Newton}=-{Gm_Am_B\over |{{\bf x_0^{B(A)}}}-{{\bf x^{A(B)}_0}}|}$ imposes the constraint:

\beqa        L_{A(B)}={G\cdot m_{A(B)}  \over 2c^2},    \label{mainb}  \eeqa as announced in the introduction.

Disregarding the self-interactions, the netto contribution of the perturbation is thus, when (\ref{mainb}) is satisfied, to add $-Gm_Am_B({1\over |{{\bf x^A_0}}-{{\bf x^B_0}}|})$ to the total energy.

If $A$ and $B$ represent identical particles, we impose that the fermionic or bosonic character of the full wave $\Psi$ is expressed at the level of the linear component only. The non-linear components behave thus like bosons and we get 

\beqa\Psi(1,2)=\Psi_L (t,{\bf x_1},{\bf x_2})\cdot ({1\over A_L})\cdot S\cdot \phi^A_{NLg}(t,{{\bf x_1}}) \phi^B_{NLg}(t,{{\bf x_2}})) \nonumber
\\ \approx e^{i\varphi_L(t,{\bf x_1},{\bf x_2})}\cdot S\cdot \phi^A_{NLg}(t,{{\bf x_1}}) \phi^B_{NLg}(t,{{\bf x_2}})),\nonumber\eeqa where
  $\phi^{A(B)}_{NLg}(t,{{\bf x_i}})=\phi_{NLg}({\bf x_i}-{\bf x_0^{A(B)}}(t))e^{-iE_g t/\hbar}$while $S$ is a bosonic symmetrization operator: 

$2S\cdot \phi^A_{NLg}(t,{{\bf x_1}}) \phi^B_{NLg}(t,{{\bf x_2}}))$=
 
 $\phi^A_{NLg}(t,{{\bf x_1}}) \phi^B_{NLg}(t,{{\bf x_2}})$+$\phi^A_{NLg}(t,{{\bf x_2}}) \phi^B_{NLg}(t,{{\bf x_1}})$. 

 The barycentres ${\bf x^{A(B)}_0}$ of the particles move according to the two particles dB-B guidance equation
  \beqa \label{guidance'b}{{\bf v_i}}_{drift}={\hbar \over m}{{\bf \bigtriangledown_i}} \varphi_L({{\bf x^{A(B)}_0}}(t),{{\bf x^{B(A)}_0}}(t),t), (i=1,2)\eeqa 
%\beqa \Psi=\Psi_L (1,2,...N)\cdot ({1\over A_L})\cdot S\cdot \pi_{j=1}^N\phi_{NLg}({\bf x^i}-{\bf x^i_0})e^{-iE_g t/\hbar},\label{appro}\eeqa
% where the barycentres ${\bf x^i_0}$ of the particles move according to the dB-B guidance equation (\ref{dBB}). 
 
In the present case, (\ref{notfreeNL}) and (\ref{guidance'b}) are dynamical equations in the 3$N$=6 dimensional configuration space (here there are $N=2$ particles). Therefore $\Psi_L (1,2)=A_Le^{i\varphi_L}$, where $A_L$ and $\varphi_L$ depend on ${\bf x^1}$ and ${\bf x^2}$  instantaneously. The system is therefore likely to exhibit a non-local \cite{EPR} behaviour in presence of entanglement \cite{bohm521}.
 %Accordingly, when, below in the text, we do not specify that we are dealing with the $i$th or $j$th particle, our results must be interpreted in the $3N$ configuration space.

We shall only consider situations where ${{\bf x^A_0}}$ and ${{\bf x^B_0}}$ do not overlap. By computations very similar to those already performed in the case of indistingusihable particles, we find that an effective gravitational interaction appears, symmetrized in ${{\bf x_1}}$ and ${{\bf x_2}}$.

\beqa \phi^G(t,{{\bf x}})/c^2=-L\Sigma_{i=1,2}({1\over |{{\bf x_i}}-{{\bf x^A_0}}|}+{1\over |{{\bf x_i}}-{{\bf x^B_0}}|})\label{sym}\eeqa

Fitting this expression with the Newton interaction provides as before the constraint $L={G\cdot m \over 2c^2}$ that we may also rewrite in the form $mc^2={G\cdot m^2 \over 2L}$ in accordance with the results announced in the introduction and in particular with the value of the energy functional (\ref{Choquard-energy}) which is of the order of $-mc^2$ in this case. Relying on previous stability analyses of the S-N equation \cite{Arriola,vanMeter,CDW}, it is worth noting that to destabilize the soliton would require a positive energy of the order of $mc^2$ which confirms our interpretation in which the soliton and the particle are the same object.

 \section{Droplets, dB-B trajectories, and gravitation\label{droplets}}

dB-B trajectories \cite{debrogliebook,bohm521,Holland} have been studied in relation with the measurement problem, but they are often considered as conceptual tools rather than empirical realities, among other reasons because the dB-B dynamics is often considered to be an ad hoc reinterpretation of the standard quantum theory. In the last decade, however, they were realized in the lab. with artificial macroscopic systems, the bouncing oil droplets or walkers  \cite{Busch,fort}, which were shown to follow dB-B like quantum trajectories.  

Our model seemingly catches some fundamental properties of droplets in the sense that it explains the appearance of an effective gravitational interaction in presence of dB-B trajectories, as they were observed directly in droplets phenomenology (and only there). 

%There remain many debatable questions that are beyond the scope of our paper; for instance have we the right to associate a macroscopic quantum wave to a droplet, or have we the right to formulate the problem in the configuration space, or yet is it equivalent to formulate the problem in the 3-dimensional space and in the configuration space. 

%Disregarding these questions for the time being, 
Of course, droplets are complicated, hydrodynamical, macroscopic systems, but in order to simplify the treatment we shall from now on identify the droplets with the self focused solitons of our model $\phi_{NL}$ and the liquid on which they are floating with  the pilot wave $\Psi_L$, which implies that the state of the system is represented by $\Psi({\bf x},t)= \Psi_L ({\bf x},t)\cdot ({\phi_{NL}^{0(A)}({\bf x},t)\over A_L({\bf x^A_0},t)}+{\phi_{NL}^{0(B)} ({\bf x},t)\over A_L({\bf x^B_0},t)})$,
where $A$ and $B$ refer to the presence of two droplets. It is then easy to repeat the same treatment as before, provided the two solitons stay away from each other. As has been reported in \cite{fort2}, stationary waves are present, resulting from the forcing frequency imposed to the container at frequency $f_0$. The stationary waves are characterized by Faraday wave lengths $\lambda_F$ in good agreement with the values computed from the dispersion relation, at the frequency $f_F$, where the Faraday frequency $f_F$ is equal to half the forcing frequency $f_0$.

In order to simulate the propagation of vibrations in the container, let us modelize, in a first step, the ``Schr\"odinger'' linear equation describing wave propagation in the container by the d' Alembert equation 
 \beqa ({1\over v^2}{\partial^2\over \partial t^2}-\Delta)\Psi_L=0,\eeqa with $v=\lambda_F\cdot f_F$. Of course d' Alembert equation is not dispersive, whereas the liquid of the container is a dispersive medium, but we shall assume in first approximation that the velocity of propagation smoothly depends on the frequency, in the spectral domain under interest. This remark matters, because, as noted in  \cite{fort2}  two frequencies play a role here, $f_0$ and $f_F=f_0/2$. 
 
 When two droplets are present we must solve an inhomogeneous d' Alembert equation with a source term $4\pi(L_A\cdot \delta({{\bf x}}-{{\bf x^A_0}})+L_B\cdot \delta ({{\bf x}}-{{\bf x^B_0}})).$
Repeating the same process as for the inhomogeneous Schr\"odinger equation, we predict the appearance of a pseudo-gravitational interaction $\phi^{PG}$ which obeys
\beqa ({1\over v^2}{\partial^2\over \partial t^2}-\Delta)\phi^{PG}=-4\pi(L_A\cdot \delta({{\bf x}}-{{\bf x^A_0}})+L_B\cdot \delta ({{\bf x}}-{{\bf x^B_0}})).\nonumber\eeqa 

Due to the forcing, we shall replace d' Alembert equation by Helmholtz equation

 \beqa ({(2\pi)^2f_0^2\over v^2}+\Delta)\phi^{PG}=(k_0^2+\Delta)\phi^{PG}=\label{waow}\\ \nonumber 4\pi(L_A\cdot \delta({{\bf x}}-{{\bf x^A_0}})+L_B\cdot \delta ({{\bf x}}-{{\bf x^B_0}})),\eeqa
%As mentioned in the conclusion (section \ref{conclu}), droplets also constitute a potential tool for confirming and/or falsifying our predictions concerning self-accelaration (attractive and repulsive as well). 

%A hidden difficulty, not to neglect, is the fact that possibly droplets ought to be treated are two-dimensional systems. Therefore,  to the contrary of the Li\'enard-Wiechert potentials, the Green functions of d' Alembert equation do not respect Huyghens principle, and have not a simple form. This of course complicates the whole treatment, but it could also open the door to a rich phenomenology and constitute an exciting field of research.

where $k_0\equiv 2\pi/\lambda_0$. %Introducing $k_F\equiv 2\pi/\lambda_F$ we get $k_0=2 k_F$.

(\ref{waow}) has been studied by us in another context \cite{vacuum}, and, in the case of a punctual source, its Green function is not $1/r$ but $cos(kr)/r$ which leads as shown in \cite{vacuum} to the appearance of attractive and repulsive gravitational zones periodically distributed among the punctual sources.

Henceforth we predict the appearance of a pseudo-gravitational field $\phi^{PG}=-v^2(L_A(cos(k_0|{{\bf x}}-{{\bf x^A_0}}|)/|{{\bf x}}-{{\bf x^A_0}}|)+L_B(cos(k_0|{{\bf x}}-{{\bf x^B_0}}|)/|{{\bf x}}-{{\bf x^B_0}}|)).$

Disregarding self-interactions, the sum of the potential energy undergone by $A$ (due to $B$) with the potential energy undergone by $B$ (due to $A$) is thus equal to 

$-v^2(cos(k_0|{{\bf x^B}}-{{\bf x^A_0}}|)/|{{\bf x^B}}-{{\bf x^A_0}}|)(M_BL_A+M_AL_B).$ Our model explains in this way the appearance of repulsive and attractive gravitation in droplets phenomenology, in connection with the presence of dB-B trajectories.

In the same vein, we explain the appearance of a pseudo-quantisation rule, self-adapting to the forcing frequency, and similar to the one observed in \cite{fort3} according to which orbital radii obey $d_n^{orb}=(n/2-\epsilon)\lambda_F$, with $n$ a positive integer, where we made use of the fact that the Faraday frequency is one halve of the forcing frequency $f_0$. In our eyes this effect is not specifically quantum, it is rather related to the very unfamiliar topology of the attractive and repulsive (!) gravitational basins (in case of circular orbits we conjecture that the aforementioned quantisation rule imposes that after one closed circular orbit each soliton keeps the Faraday phase that it possessed before the revolution). Let us now propose a toy model for quantum gravity, inspired by droplet phenomenology.

%Imposing the de Broglie-like condition that after one closed circular orbit each soliton keeps the phase that it possessed before the revolution, we may thus qualitatively predict 
%If the zones of repulsion are extended enough, we also explain the reported observation of repulsive pseudo-gravity \cite{fort} with the same argument.

\section{Minimal coupling to gravity}

Coming back to the single particle case, replacing $\Psi^{hom}_L$ by $\Psi^{hom}_L/(1-\phi^G/c^2)$ in the Schr\"odinger equation delivers the modified Schr\"odinger equation

\beqa \label{post}&&{i}\hbar\cdot  \frac{\partial }{\partial t}({ \Psi_L(t,{\bf x})\over (1-\phi^G/c^2)})= -\frac{\hbar^2}{2m}\Delta({\Psi_L(t,{\bf x})\over (1-\phi^G/c^2)})  \\ \nonumber &&+V^{L}(t,{{\bf x}})({\Psi_L(t,{\bf x})\over (1-\phi^G/c^2)})+mc^2({\Psi_L(t,{\bf x})\over (1-\phi^G/c^2)}),\eeqa

to be interpreted according to us in a perturbative approach, in terms of the small parameter $\phi^G/c^2$, replacing $1/(1-\phi^G/c^2)$ by $1+\phi^G/c^2+(\phi^G/c^2)^2+...$. There appears thus post-Newtonian corrections proportional to $ {\bf \bigtriangledown}\Psi_L(t,{{\bf x}})\cdot  {\bf \bigtriangledown} (-\phi^G(t,{{\bf x}})/c^2)$ as well as $\frac{\partial (-\phi^G(t,{{\bf x}})/c^2)}{\partial t}$ and $\phi^G/c^2$. It is not in the scope of our paper to study these corrections\footnote{These two corrections are reminiscent of the standard coupling in electro-magnetism in which we replace $i\hbar {\partial \over \partial \mu}$ by $i\hbar {\partial \over \partial \mu}-eA_\mu$. Here we replace $i\hbar {\partial \over \partial \mu}$ by $i\hbar {\partial \over \partial \mu}+i{\hbar\over c^2} {\partial \phi_G \over \partial \mu}$. Formally, the gravitational potential looks like a purely imaginary gauge field. This is not so amazing having in mind the role played by rescalings in our approach.\label{gauge}}.

We shall emit the hypothesis that the minimal coupling condition that consists of replacing $\Psi_L$ by $\Psi_L/(1-\phi^G/c^2)$ in the linear Schr\"odinger equation is general, and that $\phi^G$ is a field in ``real'', 3-dimensional space, equal to the product of the gravitational fields generated by all species of indistinguishable elementary particles. In a regime where all these contributions $\phi^k_G$ are small compared to $c^2$, we find that  $\pi_{ k}(1-\phi^k_G/c^2)\approx (1-\sum_k \phi^k_G/c^2)$, which expresses the universal character of gravitation.

If we had chosen to begin with to describe the system with the relativistic Klein-Gordon equation (that we write in the free case here for simplicity but the result is still valid in the presence of external potentials e.g. electro-magnetic potentials)\beqa (c^2({1\over c^2}{\partial^2\over \partial t^2}-\Delta)-{m^2 c^2\over \hbar^2})\Psi_L=0\eeqa then, by a treatment similar to the one performed in the previous sections we would have found finally that the gravitational potential obeys the inhomogeneous d' Alembert equation:

\beqa-c^2({1\over c^2}{\partial^2\over \partial t^2}-\Delta)\phi_G=4\pi G \rho.\label{inhomDAL}\eeqa We find so that all gravitational waves move at the (same) speed of light in vacuum.

Beginning with the Dirac equation \beqa \beta i\hbar \partial_t {\bf \Psi}(t,{\bf x})-\beta{\bf \alpha}c{\hbar\over i}{\bf \bigtriangledown}{\bf \Psi}(t,{\bf x})=mc^2{\bf \Psi}(t,{\bf x})+V_{NL}{\bf \Psi}(t,{\bf x}),\nonumber\eeqa
where $\alpha$ and $\beta$ represent Dirac matrices and imposing the ansatz ${\bf \Psi}={\bf \Psi_L}(t,{{\bf x}})\cdot \phi_{NL}(t,{{\bf x}})\label{ansatz}$ where ${\bf \Psi_L}(t,{{\bf x}})$ a Dirac 4-spinor and $\phi_{NL}(t,{{\bf x}})$ is a Lorentz scalar moving at the velocity ${\bf v}_{DiracdB-B}={\bf  \Psi_L}(t,{\bf x})^\dagger{\bf \alpha}{\bf  \Psi_L}(t,{\bf x})/{\bf  \Psi_L}(t,{\bf x})^\dagger{\bf  \Psi_L}(t,{\bf x})$, in accordance with the dB-B guidance equation \cite{Takaba,Holland}, we find again by similar methods as those of previous sections the inhomogeneous Dirac equation
\beqa i\hbar \partial_t {\bf \Psi_L}(t,{\bf x})-{\bf \alpha}c{\hbar\over i}{\bf \bigtriangledown}{\bf \Psi_L}(t,{\bf x})-mc^2\beta {\bf \Psi_L}(t,{\bf x})=\nonumber \\-{\hbar\over i}{\bf \alpha}c({\bf \Psi_L}(t,{\bf x})-{\bf \Psi_L}(t,{\bf x_0}))  {{\bf  \bigtriangledown}\phi_{NLg}(t,{\bf x})\over \phi_{NLg}(t,{\bf x})}\label{Diracsource}\eeqa  Multiplying (\ref{Diracsource}) by $( i\hbar \partial_t +{\bf \alpha}c{\hbar\over i}{\bf  \bigtriangledown}+mc^2\beta)$ we get $-\hbar^2c^2({1\over c^2}({\partial^2\over \partial t^2}-\Delta)+{m^2 c^2\over \hbar^2}) {\bf \Psi_L}(t,{\bf x})=( i\hbar \partial_t +{\bf \alpha}c{\hbar\over i}{\bf  \bigtriangledown}+m\beta){(-)\hbar\over i}{\bf \alpha}c({\bf \Psi_L}(t,{\bf x})-{\bf \Psi_L}(t,{\bf x_0})) {{\bf  \bigtriangledown}\phi_{NLg}(t,{\bf x})\over \phi_{NLg}(t,{\bf x})}$.

Applying the same reasonings as in the previous section, making use of (\ref{mainb}), and $\alpha_k^2=1$, still in the limit of slowly moving bodies, we find (\ref{inhomDAL}). Together with 

$ i\hbar \partial_t {{\bf  \Psi_L}(t,{\bf x})\over (1-\phi_G/c^2)}-{\bf \alpha}c{\hbar\over i}{\bf \bigtriangledown}{{\bf  \Psi_L}(t,{\bf x})\over (1-\phi_G/c^2)}-mc^2\beta {{\bf  \Psi_L}(t,{\bf x})\over (1-\phi_G/c^2)}=0$,  

 and the Dirac-dB-B guidance relation ${\bf v}_{DiracdB-B}={\bf  \Psi_L}(t,{\bf x})^\dagger{\bf \alpha}{\bf  \Psi_L}(t,{\bf x})/{\bf  \Psi_L}(t,{\bf x})^\dagger{\bf  \Psi_L}(t,{\bf x})$ our model is self-consistent.

\section{Conclusions} Treating particles/droplets as incompressible solitons which obey dB-B dynamics\footnote{In \cite{old} we explain how we were brought to impose (\ref{approsingle},\ref{bari},\ref{dBB}) that in the present paper we treat as a postulate. Actually in \cite{old} we studied the feedback of the pilot wave $\Psi_L$ on the soliton or particle $\phi_{NL}$, postulating to begin with that (\ref{notfreeNL},\ref{selfclas}) is satisfied and also that $\Psi=\Psi_L\cdot \phi_{NL}$, while $\Psi_L$ is a solution of the homogeneous Schr\"odinger equation; we argued that this feedback results in conditions (\ref{approsingle},\ref{bari},\ref{dBB}) in the appropriate regime. \cite{old} is complementary to the present paper in the sense that here we study the feedback of the solitons on the pilot wave.}, the spread of which is compensated by a self focusing non-linear interaction, we predicted the appearance of an effective gravitation. In the case of elementary particles, our model makes it possible, fitting the effective gravitation with Newtonian gravitation to estimate the extent of the particle (\ref{mainb}). In the case of droplets, it delivers precise predictions about the pseudo-gravitational interaction. Our model is quite unorthodox. We found our main inspiration in old fashioned concepts such as the Poincar\'e pressure \cite{Poincar} or de Broglie's double solution program \cite{debrogliebook},  and we did not make any kind of reference to curved space-time. This is maybe an open door for quantum gravity, maybe not. At this level, we simply do not know. In order to falsify our model it suffices in principle to test its post-Newtonian predictions, which is beyond the scope of our paper. According to us, alternative approaches to gravity could maybe help to understand apparent modifications of Newton's equation in the solar system \cite{Reynaud} and in galactic rotation curves as well \cite{vacuum} but this is another story. In the meanwhile, our study suggests that dB-B trajectories and effective gravitation could be observed with collective self-collapsed assemblies of cold atoms \cite{Kaiser}, which has never been done so far. In any case, a virtue of our model  is that it has some retrodictive power in the sense that it explains the appearance, in presence of dB-B trajectories, of a pseudo-gravitational interaction in droplets phenomenology, attractive and repulsive as well. Moreover we predict that the interaction scales like $v^2(L_AM_B+L_BM_A)cos (k_0r_{AB})/r_{AB}$, which can possibly be tested in the lab..    There are two unspecified parameters in our model, at this level, which are the exact ratios between $L_{A(B)}$ and the sizes of the droplets. We expect those to be of the order of unity however, and they can easily be evaluated experimentally.

\section*{Acknowledgments}
This work benefitted from the support of grants 21326 and 60230 from the John Templeton Foundation and the FQXI project FQXi-RFP-1506.
  Thanks to C. Champenois and R. Willox for kind comments.

\bibliographystyle{plain}

\end{document}